\documentclass[twocolumn,showpacs,preprintnumbers,amsmath,amssymb,aps,pre]{revtex4}

\usepackage{latexsym}
\usepackage{dcolumn}% Align table columns on decimal point
\usepackage{supertabular,multirow}
\usepackage{epsfig}
\usepackage{bm}% bold math
\usepackage{subfigure}
\usepackage{amsmath}
\usepackage{graphicx}
\usepackage{color}

\newcommand{\refeq}[1]{Eq. \ref{#1}}
\newcommand{\Ca}{\mbox{\it Ca}}

\newcommand{\Rr}{\mbox{\it R}}
\newcommand{\Sr}{\mbox{\it S}}

\newcommand{\rhat}{{\bf \hat r}}
\newcommand{\zhat}{{\bf \hat z}}
\newcommand{\bx}{{\bf x}}

\newcommand{\out}{{\mathrm{ex}}}
\newcommand{\ins}{{\mathrm{in}}}

\newcommand{\el}{{\mathrm{el}}}

\newcommand{\bt}{{\bf t}}
\newcommand{\visrat}{{\lambda}}

\newcommand{\eps}{{\varepsilon}}

\newcommand{\bE}{{\bf{E}}}

\DeclareMathOperator{\arccot}{arccot}

\newcommand{\sigm}{{\sigma}}

\begin{document}

\author{M. Ouriemi}
\affiliation{IFPEN, Solaize, BP 3 69360, France}
\author{P. M. Vlahovska}
\affiliation{School of Engineering, Brown University, Providence, RI, 02912, USA}
\email{petia_vlahovska@brown.edu}

 \title{ Electrohydrodynamic deformation and rotation of a particle-coated drop}

\date{\today}

\begin{abstract}
A dielectric  drop suspended in conducting liquid and  subjected to  an uniform electric field  deforms into an ellipsoid whose major axis  is either perpendicular  or tilted (due to  Quincke rotation effect) relative to the applied field. We experimentally study the effect of surface-adsorbed colloidal  particles on these classic electrohydrodynamic phenomena. 
We observe that  at high surface coverage ($>90\%$),  the electrohydrodynamic flow is suppressed,
oblate drop deformation is enhanced, and the threshold for tilt is decreased compared to the particle--free drop.  The deformation data are well explained by a capsule model, which assumes that the particle monolayer acts as an elastic interface. The reduction of the threshold field for rotation is  likely related to drop asphericity.
\end{abstract}
\maketitle

\section{Introduction}

\label{sec:intro}

{ {Electric fields provide a versatile means to control small-scale fluid and particle dynamics, e.g., electrohydrodynamic instabilities for pattern formation in thin polymer films \cite{ Wu-Russel, Wu-Pease-Russel:2005, Kumar:2009} or particle suspensions \cite{Lin:2014}, electrohydrodynamic atomization to produce micro- and nano-particles \cite{delaMora, Collins:2008, Xie:2015, Farook:2009, Mahaingam:2014}, drop and vesicle manipulation \cite{Lecuyer-Ristenpart:2008, Vigo:2010}, and colloidal assembly  \cite{Ristenpart:2008, Velev:2009, Prieve:2010, Blaaderen:2013, Aranson:2013, Dobnikar:2013,  Ristenpart:2015}. }}
Fluid interfaces provide additional functionality opening new routes for the  bottom-up fabrication of novel structurally 
complex  materials \citep{Lee:2013, Furst:2011, Cavallaro:2011, Cui:2013}.  
Recent works \cite{Dommersnes, Ouriemi:2014, Rozynek:2014, Rozynek:2014b} find that
micro--particles constrained on a drop surface can form various structures in the presence of applied uniform electric field.  The underlying mechanisms are still under investigation but a major driving force in this system is the flow created by the electric shear stresses due to accumulation of charges at the interface. 

\begin{figure}[h]
\centerline{\includegraphics[width=2.75in]{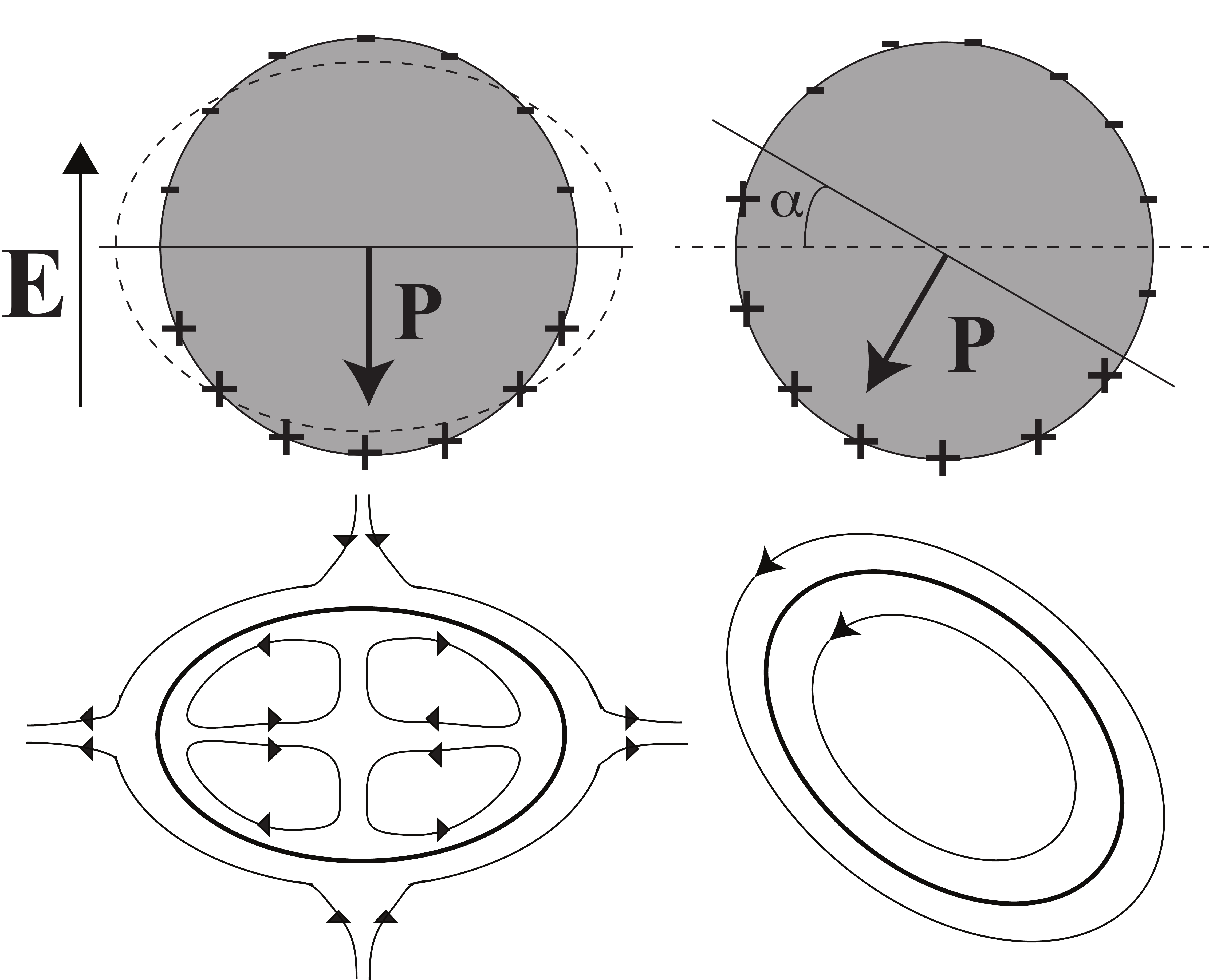}}
\caption{ \footnotesize Drop polarization and electrohydrodynamic flow streamlines for $RS<1$ in applied uniform DC electric field. 
 (a)  Weak fields with $E_0<E_Q$,  where $E_Q$ is given by \refeq{quinckeW},  induce pure straining flow and axisymmetric oblate deformation. (b) In strong fields, $E_0>E_Q$ constant torque  is induced by the misaligned dipole,  the flow acquires a  rotational component, and the drop is tilted with respect to the applied field direction. }
  \label{fig1}
  \end{figure}

A particle-free drop placed in an electric field polarizes 
because of the 
mismatch of  the bulk fluids electrical conductivity,  $\sigm$, and  dielectric constant,  $\eps$
\begin{equation}
\Rr=\frac{\sigm_\ins}{\sigm_\out}\,,\quad \Sr=\frac{\eps_\out}{\eps_\ins}\,.
\label{ParameterRatios}
\end{equation}
Upon application of an electric field, mobile charges
 carried by conduction  accumulate at the boundary (even though the net charge on the interface remains zero), see    Figure \ref{fig1}.
  For a sphere placed in a uniform electric field with magnitude $E_0$, the induced surface free charge distribution  
increases as  \cite{Lanauze:2013, JonesTB}
\begin{equation}
\begin{split}
Q=&3\eps_\ins E_0 \frac{1-RS}{R+2}(1- e^{-t/t_{mw}})\sin\theta \,\\
 t_{mw}=&\frac{\eps_\ins+2 \eps_\out}{\sigm_\ins+2\sigm_\out}\,.
 \end{split}
\end{equation}
where $\theta$ is the angle with the applied field direction and $t_{mw}$ is the Maxwell-Wagner polarization time.    The subscripts $``\ins "$ and $``\out "$ denote the values for drop and suspending medium, respectively. 
Thus, the polarity of the induced charge 
 is determined by the product of $R$ and $S$, which compares the conduction response of the media \cite{Melcher-Taylor:1969, Saville:1997}.
If $RS<1$, the conduction in the drop is slower than  the suspending liquid.  As a result, the interface charge distribution is dominated by charges brought from the suspending medium and the drop dipole moment is oriented opposite to the applied electric field direction (note that in the opposite case,   $RS>1$, the 
particle dipole is reversed and it is aligned with the electric field.)

The electric field acting on the induced surface charges creates a tangential stress, $\tau_{r\theta}=QE_\theta$.
In the case of a simple fluid-fluid interface such as oil-water this shear stress  can only be balanced by viscous stresses due to fluid flow. 
In weak fields, for which the electric stresses $\eps_\out E_0^2$ are smaller 
than the capillary stress due to surface tension $\gamma/a$ so that drop shape remains nearly spherical  \begin{equation}
\Ca=
 \frac{a\eps_{\out}E_0^2}{\gamma}\ll 1 \,,
\end{equation}
where  $a$ is the drop radius and $\gamma$ is the interfacial tension, at steady state the fluid  undergoes axisymmetric straining flow about the drop~\cite{Taylor:1966}, see    Figure \ref{fig1}.(a), with surface velocity 
\begin{equation}
\label{svel}
u_\theta=\frac{2 \tau_{r\theta}}{1+\visrat}=\frac{9\eps_\ins E_0^2(1-RS)}{(1+\visrat) (R+2)^2} \sin(2\theta)\,.
\end{equation}
where  $ \visrat={ \mu_\ins}/{ \mu_\out}$ is the viscosity ratio.
{ {The  charge distribution corresponding to $RS<1$, illustrated in    Figure \ref{fig1}.(a), results in surface fluid motion from the pole to the equator and the equator is a  stagnation line.}}
 Drop deformation resulting from the action of the electric stresses is  either oblate  or prolate ellipsoid \cite{Taylor:1966} 
\begin{equation}
\label{Def_param}
\begin{split}
D=&\frac{d_{||}-d_\perp}{d_{||}+d_\perp}=\frac{9\Ca }{16 S }\Phi(R,S,\lambda)\,,\\\
 \Phi=&\frac{1}{(2+R)^2}\left[S(R^2+1)-2+3(RS-1)\frac{3\lambda+2}{5\lambda+5}\right]
\end{split}
\end{equation}
where $d_{||}$ and $d_\perp$ are the spheroid axes parallel and perpendicular to the direction of the applied 
electric field.

Recently it was observed that  a nonaxisymmetric rotational flow may appear for drops with $RS<1$, see   Figure \ref{fig1}.(b). The drop can assume steady tilted orientation with respect to the electric field direction \cite{Ha:2000a, Salipante-Vlahovska:2010} or oscillate \cite{Sato:2006, Salipante-Vlahovska:2013}. These behaviors have been linked to the Quincke rotation phenomenon\cite{Quincke:1896, Jones:1984,Turcu:1987, Lemaire:2002}, which is an instability arising from the unfavorable orientation of the induced dipole (for $RS<1$ the dipole direction is opposite to the applied electric field and ``wants'' to flip).
The complete flip of the dipole is prevented by charge supply from the bulk:  the induced surface charge (dipole)  rotates with the particle, but at the same time the suspending fluid  recharges the interface. The balance between charge convection by rotation and supply by conduction from the bulk results in a steady misaligned torque and continuous spinning in the case of a rigid sphere, the so called Quincke rotation. The rotation rate $\omega$ and the steady oblique dipole orientation, characterized by the angle $\alpha$ (illustrated in    Figure \ref{fig1}.(b)) are
\begin{equation}
\label{quinckeAngle}
 \omega=\textstyle{\frac{1}{t_{mw}}\sqrt{\frac{E^2}{E_Q^2}-1}\,,\quad  \alpha=\arccot \left[\left(\omega t_{mw}\right)^{-1}\right]\,}
 \end{equation}
where $E_Q$ is the threshold electric field above which the rotation occurs and  $t_{mw}$ is the Maxwell-Wagner polarization time
\begin{equation}
\label{quinckeW}
\begin{split}
E_Q^2=\frac{2\sigm_\out \mu_\out \left(\Rr+2\right)^2}{3\eps_\out \eps_\ins(1-\Rr\Sr)}\,,
  \end{split}
\end{equation}
$\Rr$ and $\Sr$ are the conductivity and permittivity ratios defined by  \refeq{ParameterRatios}.
In the case of drops, the electric torque generates rotational fluid motion \cite{Ha:2000a, Salipante-Vlahovska:2010}, in addition to the already present straining flow.  The resulting linear flow causes the drop to deform into a general ellipsoid whose major axis is misaligned with the applied electric field.The steady oblique orientation and deformation of the drop were experimentally and theoretically investigated by our  group \cite{Salipante-Vlahovska:2010, He:2013, Salipante-Vlahovska:2013}.  

Given the complex behavior of a drop with a simple fluid interface, a question naturally arises:  how does surface modification changes drop electrohydrodynamics? We studied experimentally  a microparticle-coated drop with $RS<1$ in  the Taylor regime (below the threshold for Quincke rotation) and at low to moderate surface coverages (below 50\%) \cite{Ouriemi:2014}. Particles initially randomly distributed at the surface formed a ``belt" around the equator as the drop deforms into an oblate shape.  This is an expected consequence of the straining electrohydrodynamic flow. In stronger DC fields,  the dynamics of these ``armored'' drops becomes more complex  \cite{Dommersnes, Ouriemi:2014} and  depends on the particles
characteristics. Belts formed by low polarizability particles break into a sequence of counter-rotating vortices of particles. When dipole--dipole attraction becomes
strong the particle chain and the drop experiences a prolate deformation and
tip-streaming occur with ejection of particles. For non-spherical conductive particles,
we have observed drop ``kayaking'': its major axis precessing around the applied field direction. 

In this work we investigate drops at high particle-coverage. The particle--covered drop is modeled as a capsule and the experimentally measured drop deformation, $D$, as a function of field strength, $\Ca$, is used to find the shear elasticity of particle monolayers. We also investigate the effect os particle coverage on the onset for  Quincke rotation, and we find that high surface coverage of particles drastically reduces the critical fields strength of rotation.

\section{Experimental section}

\label{sec:exp}

\subsection{Materials}
\label{subsec:Materials}
The suspending fluid is castor oil  (Alfa Aesar) with  viscosity $\mu_{\out} = 0.69$ Pa.s, dielectric constant  $\eps_{\out} = 4.6\eps_0$, where $\eps_0$ is the vacuum permittivity, conductivity  $\sigma_{\out} = 3.8\times10^{-11}$S/m, and density $\rho_{\out} = 962\;kg/m^{-3}$. The drop fluid is silicon oil  with $\mu_{\ins} = 0.05$ Pa.s, $\eps_{\ins} = 2.8\eps_0$, $\sigma_{\ins} = 2.4\times10^{-12}$S/m, $\rho_{\ins} = 963.5\;kg/m^{-3}$ (UCT). The surface tension between the castor oil and the Silicone oil is $4.5mN/m$ \cite{Salipante-Vlahovska:2010}. The permittivity and the conductivity were measured using a dielectric constant meter and a conductivity meter from Scientifica. The viscosity and density are specified by the company-provider. The characteristics of the particles are summarized in Table \ref{tab:part}.
\begin{widetext}
\begin{table}[h]
  \begin{center}
\def~{\hphantom{0}}
  \begin{tabular}{lccccccc}
      Shape & Type & density $\rho_p$ ($kg/m^3$) & radius $r$ ($\mu m$)& Conductivity & Supplier  \\[3pt]
         irregular & Aluminum (Al)& 2600 &$ 1.5, \; 12 $ & $++$ & Atlantic equipment  \\
         sphere & glass (G)& 2200 &  $ 3.5, \; 8.5$ & $+$ & Corpuscular/Cospheric\\
      sphere & polyethylene (Pe) & 1000 &  50 & $-$ & Cospheric  \\
      \end{tabular}
  \caption{\footnotesize Particles characteristics. The symbol $+$ indicates that the particles are slightly more conducting that the fluids, $++$ highly more conducting, and $-$ less conducting. 
  }
  \label{tab:part}
  \end{center}
\end{table}
\end{widetext}

\subsection{Methods}

\begin{figure}[h]
\centerline{\includegraphics[width=3in]{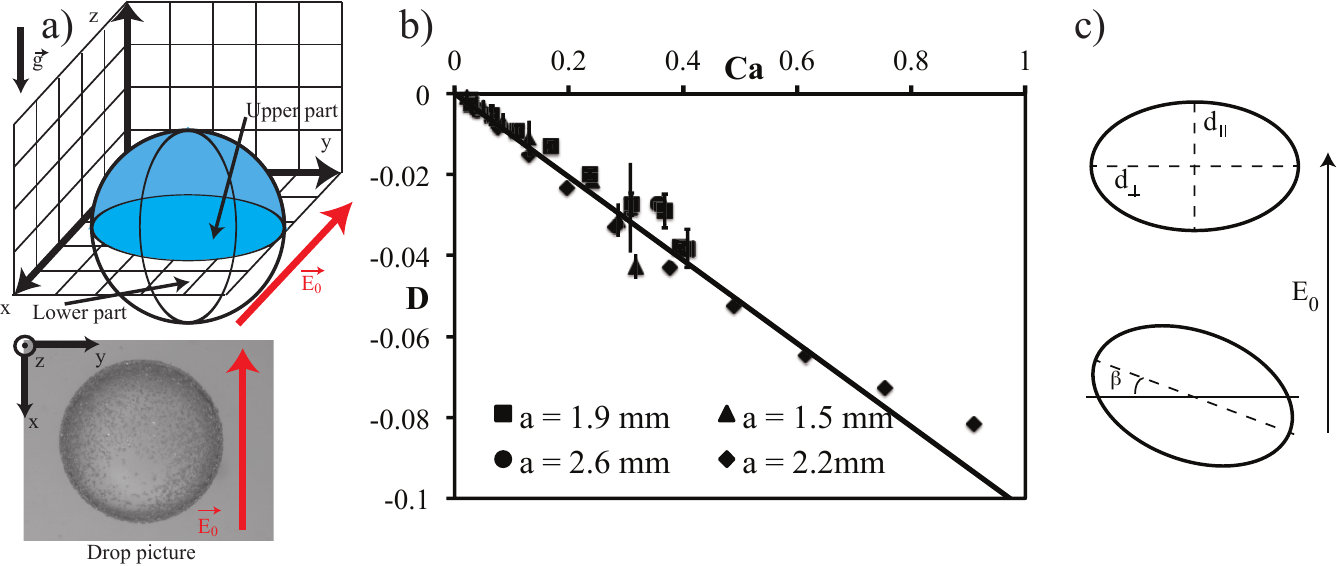}}
\caption{\footnotesize{(a) Drop observation for the experimental measurements. (b) $D(\Ca)$ for a particle--free drop is used to validate the set-up. The symbols are experimental data and the line is the Taylor's theory \refeq{Def_param}. (c) Definition of drop deformation and orientation characteristics: the spheroid axes $d_{||}$ and $d_\perp$  
% parallel and perpendicular to the direction of the applied electric field 
and the  tilt $\beta$.}}
\label{fig2}
\end{figure}

 The experimental set-up has the same design as the one  used in \cite{Ouriemi:2014}. 
A uniform DC electric field is created in a parallel-plate chamber made of  two 7.6cm by 10.5 cm brass  electrodes separated by a gap  of 4.5cm. Fields up to 16 $kV/cm$ are generated using a voltage amplifier connected to a DC Power supply.  The chamber is filled with castor oil. A drop of silicon oil with suspended micro-particles is injected in the middle of the chamber. The particles are driven to the interface by the  application of an electric pulse the duration and strength of which depend on the type of micro-particles;  after the field is turned off the particles remain  trapped at the interface  as the thermal energy is negligible compared to the energy required to remove the particle from the interface.The drop is then manually moved around in order to randomize the particle distribution on the drop surface. In the actual experiment, electric field is applied and drop behavior is recorded for about 3 minutes. On this time scale drop sedimentation is negligible. After each recording, the electrical field is turned off  and the drop is moved back to its initial position. This action removes any particle structures.  The experiment is repeated for a different electric field strength.     

The drop is observed from a direction perpendicular to the field. The top view of the drop is recorded every 0.2s,  which is a compromise between memory limitation of the software and the length of the recording.
 The images are post-processed with ImageJ and Matlab to extract the drop diameter $d$, { {the deformation parameter $D=(d_{||}-d_\perp)/(d_{||}+d_\perp)$}} and the angle $(\pi/2 -\beta)$ between the drop major axis  and the applied field direction, see    Figure \ref{fig2} for definitions. The precision of the measurements  is set by the resolution of the pictures; for a drop of radius $a = 2 mm$ it is around 0.0055 mm/px leading to an absolute error in the deformation (D) of 0.0045, which decreases for smaller drops. The surface concentration of particles, $\varphi$,  is defined as the percentage of the drop surface covered by particles (including the space between particles) once the particles are brought together. We measure $\varphi$ at very low electrical fields, $E_0 <60kV/m $, to minimize  particle compaction. The experimental procedure and the material properties are validated by comparing drop deformation with the Taylor model for small deformations\cite{Taylor:1966} given by \refeq{Def_param} .

\section{Results}

\begin{figure}[h]
\includegraphics[width=3in]{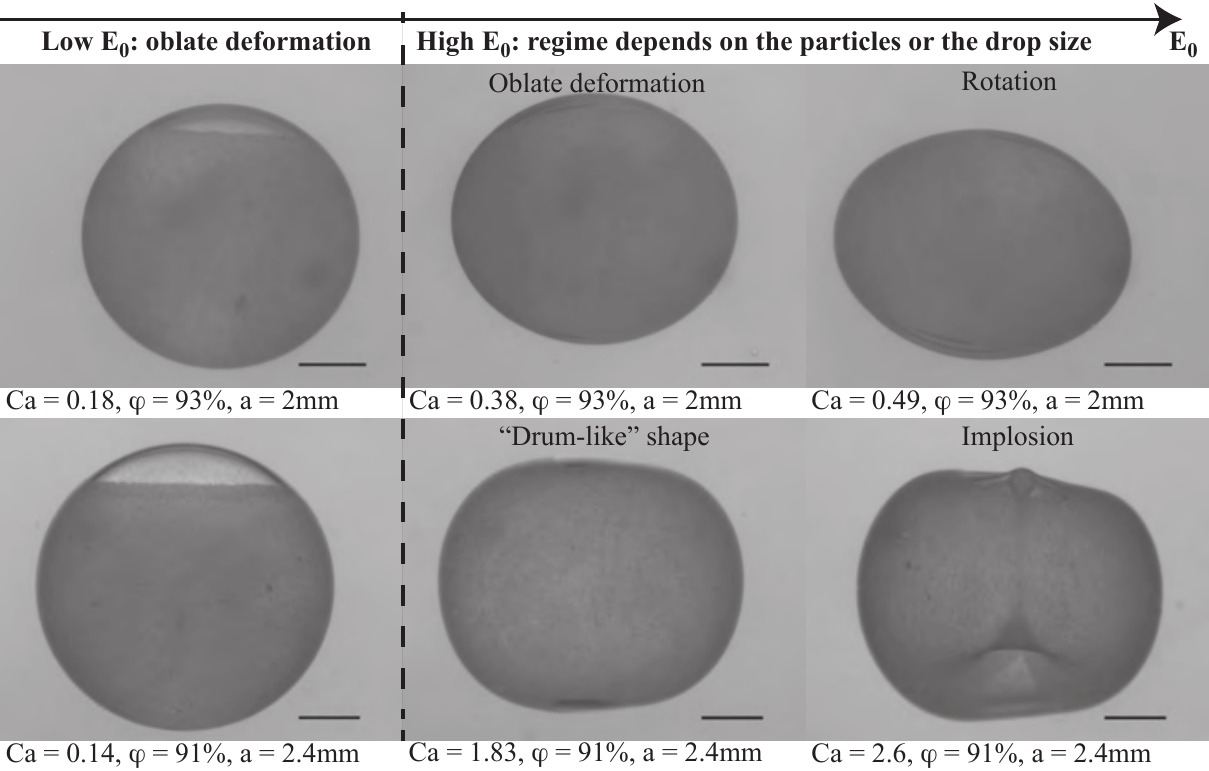}
\caption{\footnotesize{Typical drop behaviors at high surface coverage of glass spheres ( with radius $r=8.5\mu m$). { { The scale bar corresponds to 1 mm.}}}}
\label{fig3}
\end{figure}

   Figure \ref{fig3} illustrates the typical behaviors of drops with $\varphi\sim 90\%$ surface coverage. In weak fields,  the drop always deforms into an oblate spheroid and the particles accumulate at the equator forming a ``belt''.  At high coverages, the belt is very wide and  only a small region near the poles remains particle free.  In stronger fields, the oblate deformation increases and the drops may adopt peculiar  ``drum-like'' shapes. In even stronger fields,  the drops may start rotating or implode. 

\subsection{Weak DC electric fields: increased oblate deformation}
\begin{figure}[h]
 \centerline{\includegraphics[width=3in]{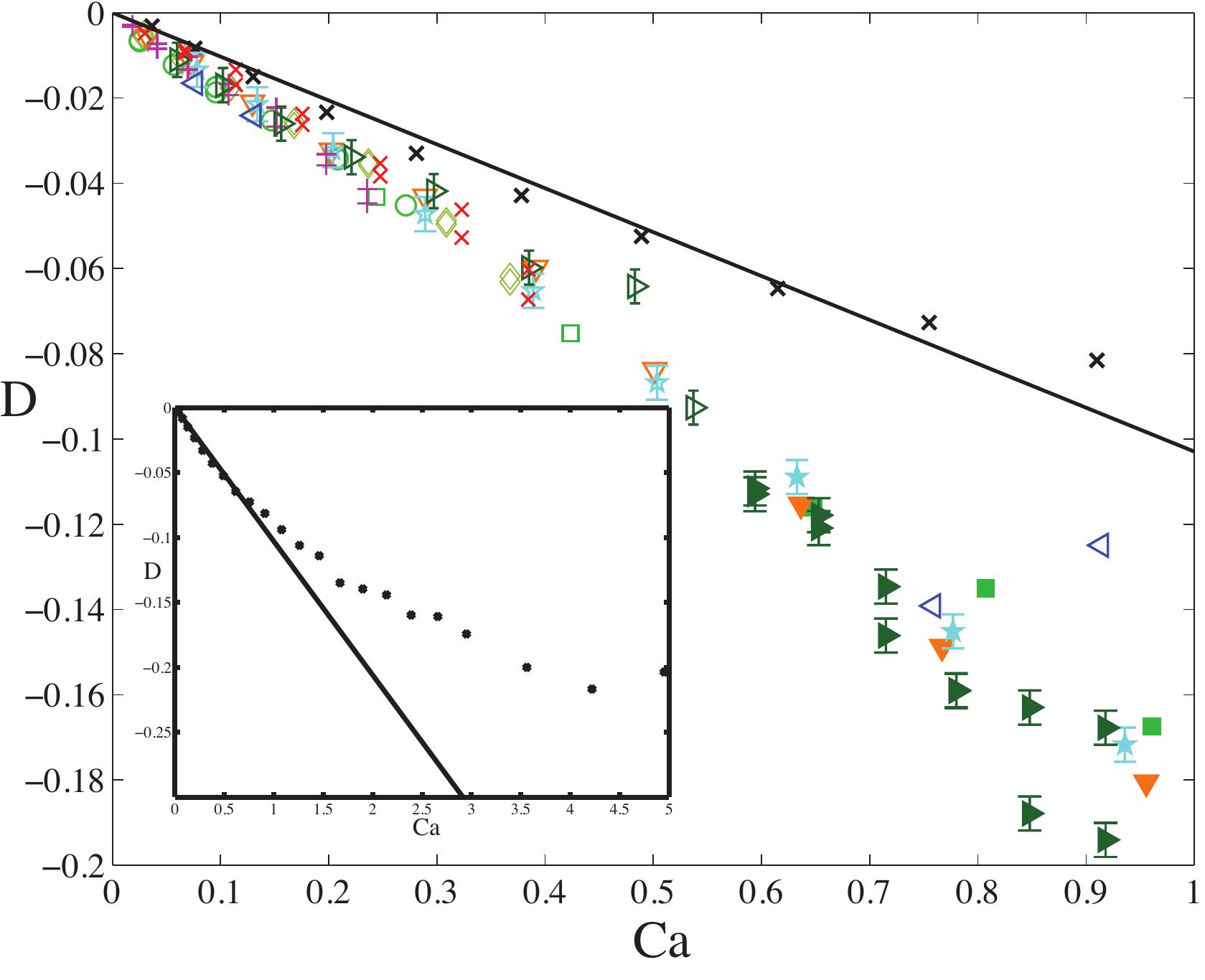}}
\caption{\footnotesize{
Drop deformation $D$ as a function of the dimensionless  field strength $\Ca$. The symbol $\times$ corresponds to a ``clean'' (particle-free)  drop with radius $a$ = 2.2mm, the symbols $\diamond$, $\square$,
 $\circ$ and $\bigtriangleup$ denote drops covered with Al ($r=12\mu m$) with  $\varphi = 65\%$, $88\%$, $93 \%$ and $96\%$ and $a$ = 1.9mm, 1.8mm, 1.7mm and 2.2mm, respectively. The symbol $\star$ corresponds to a drop covered with  Al ($r=1.5\mu m$) with $\varphi = 80\%$ and $a$ = 2.3mm, the symbol $\triangleleft$ corresponds to a drop covered with Pe ($\varphi = 78\%$, a = 2.2mm), the symbols $+$, and light $\times$ denote drops covered with G ($r=8.5\mu m$) with  $\varphi = 100\%$ and $93\%$ and a = 1.2mm, and 2mm, the symbol $\triangleright$ corresponds to a drop covered with G ($r=3.5\mu m$) with $\varphi = 100\%$ and a = 1.7mm. The solid line corresponds to the Taylor prediction \refeq{Def_param} for viscosity ratio $\lambda = 0.07$. The inset shows the deformation of a particle--free drop in a wider range of $\Ca$. Filled symbols denote drops in the electrorotation (tilted) regime. For clarity,  error bars are shown only for some drops as they are similar for all the measurements.}}
\label{fig4}
\end{figure}

   Figure \ref{fig4} shows that  particle--covered drops undergo larger deformation compared to uncoated (``clean'') drops at the same field strength. Moreover, the magnitude of the deformation is insensitive to particle type, size, and coverage (above $\varphi=65\%$). 

\begin{figure}[h]
\includegraphics[width=3in]{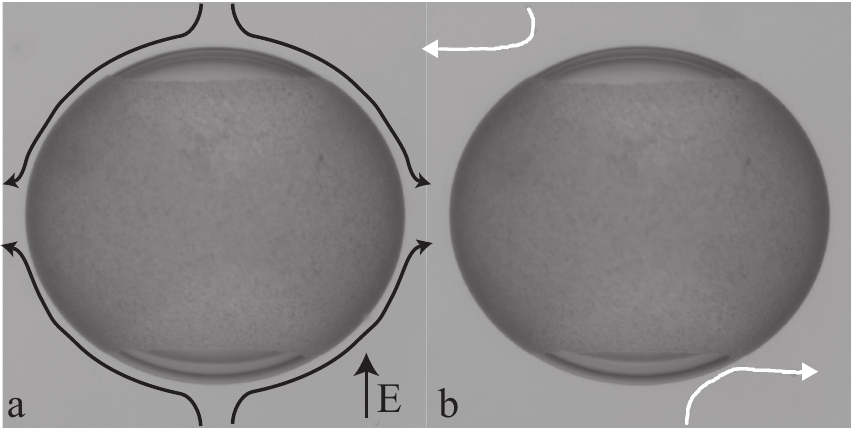}
\caption{\footnotesize{Streamlines of the flow around a drop. a) Theoretical flow around a clean drop. b) Trajectories measured around a glass covered drop }}
\label{fig:traj}
\end{figure}

Once the drop reaches a steady oblate deformation, the particles at the surface do not move, which suggests the absence of electrohydrodynamic flow. To test for the presence of flow, few large polyethylene spheres (Pe) were used as tracers in the surrounding fluids furnishing a qualitative visualization of the flow around the drop.    Figure \ref{fig:traj}  illustrates the streamlines for the flow  around a ``clean'' drop and the measured trajectories  of two particles around a drop covered with glass particles. For high surface coverage, 
the traces flow  from the poles to the belt boundary. Tracers in the vicinity of the surface covered by particles are still. Thus, the presence of particles suppress the electrohydrodynamic flow and  limits it  to  the poles region (`clean' portion of the drop).

\subsection{Stronger DC electrical fields: drop tilt}

\begin{figure}[t]
\includegraphics[width=3.25in]{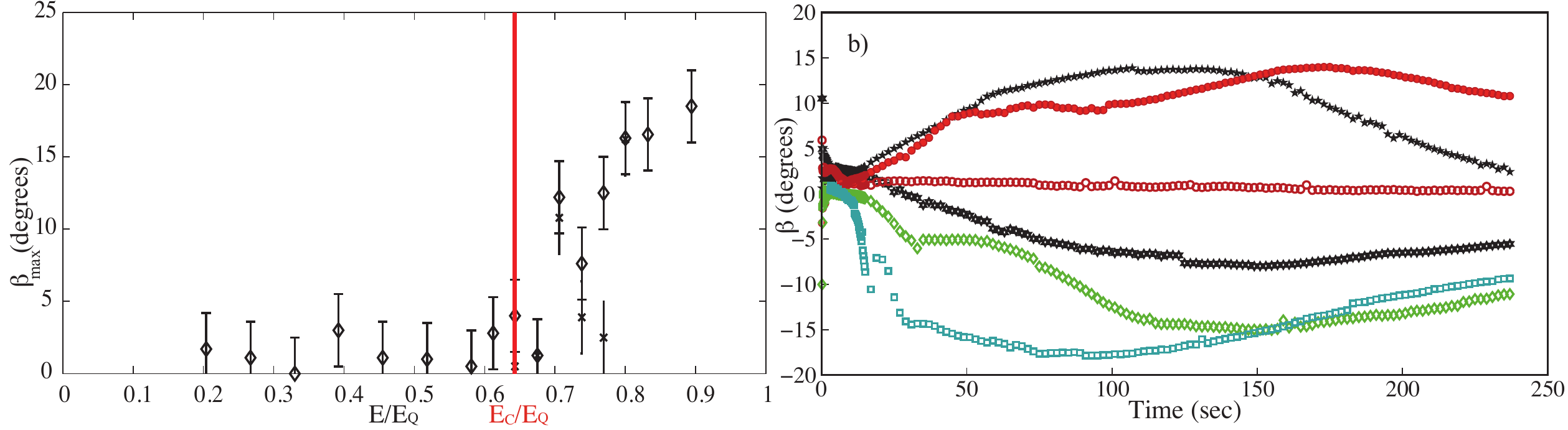}
\caption{\footnotesize{(a) - Maximum inclination $\beta_{max}$ measured over 4 minutes as a function of $E/E_{Q}$} for a drop covered with G spheres ($r = 3.5\mu m$). The symbol $+$ represents the second measurement realized for some $E/E_{Q}$. b- Examples of some temporal evolution of $\beta$. The symbol $\square$ corresponds to $E/E_{Q} = 0.89$, the symbol $\star$ to $E/E_{Q} = 0.70$, the symbol $\circ$ to $E/E_{Q} = 0.77$, the symbol $\diamond$ to $E/E_{Q} = 0.80$ and the line to $E/E_{Q} = 0.45$.The filled symbols represent the second measurement realized for some $E/E_{Q}$.}
\label{fig:beta_glass}
\end{figure}

Particle--covered drops also exhibit Quincke-like (electrorotation) behavior.
We find that  the inclination angle does increase with field strength, but also exhibits unsteady behavior, see    Figure \ref{fig:beta_glass}. The appearance of  nonzero tilt $\beta$ (steady or unsteady) 
 is defined as the threshold for rotation.

   Figure \ref{fig:beta_glass}.(b) represents the time evolution of the tilt angle $\beta$ (see  Figure \ref{fig1} for the definition) measured for a drop covered of small glass spheres for different electrical fields.  For low electrical fields, $\beta$ 
remains nearly zero, i.e., there is no tilt.
 The small angle is due to errors in the position of the electrical chamber relative to the camera. 
 In stronger field the drop tilts.
 However, in contrast to  `clean' drops which exhibit stationary inclination, the inclination ($\beta$) varies with time, see    Figure \ref{fig:beta_glass}.(b), similarly to the wobbling observed for drops with a low coverage of aluminum particles  \cite{Ouriemi:2014}. { { time scale of the transients is comparable to the duration of the observation, so we can not conclusively state that the behavior is periodic (e.g., oscillations). }}   Figure \ref{fig:beta_glass}.(a) represents the maximum inclination $\beta_{max}$ measured as a function of the  electrical fields scaled by the critical value for Quincke rotation ($E/E_{Q}$). Filled symbols correspond to repeated experiments. Below the rotation threshold, $E_{c}/E_{Q}$, $\beta_{max}$ is within the experimental error; { {$E_{c}$ denotes the critical electric field for the onset of electrorotation in our system (particle-coated drop)}}.
 Above $E_{c}$, $\beta_{max}$ increases with field strength $E$. Repeated measurements show that for a given $E$, different inclinations are possible. Finally we observe sometimes that the a drop can spin while having no tilt.

\begin{figure}[h]
\includegraphics[width=3.25in]{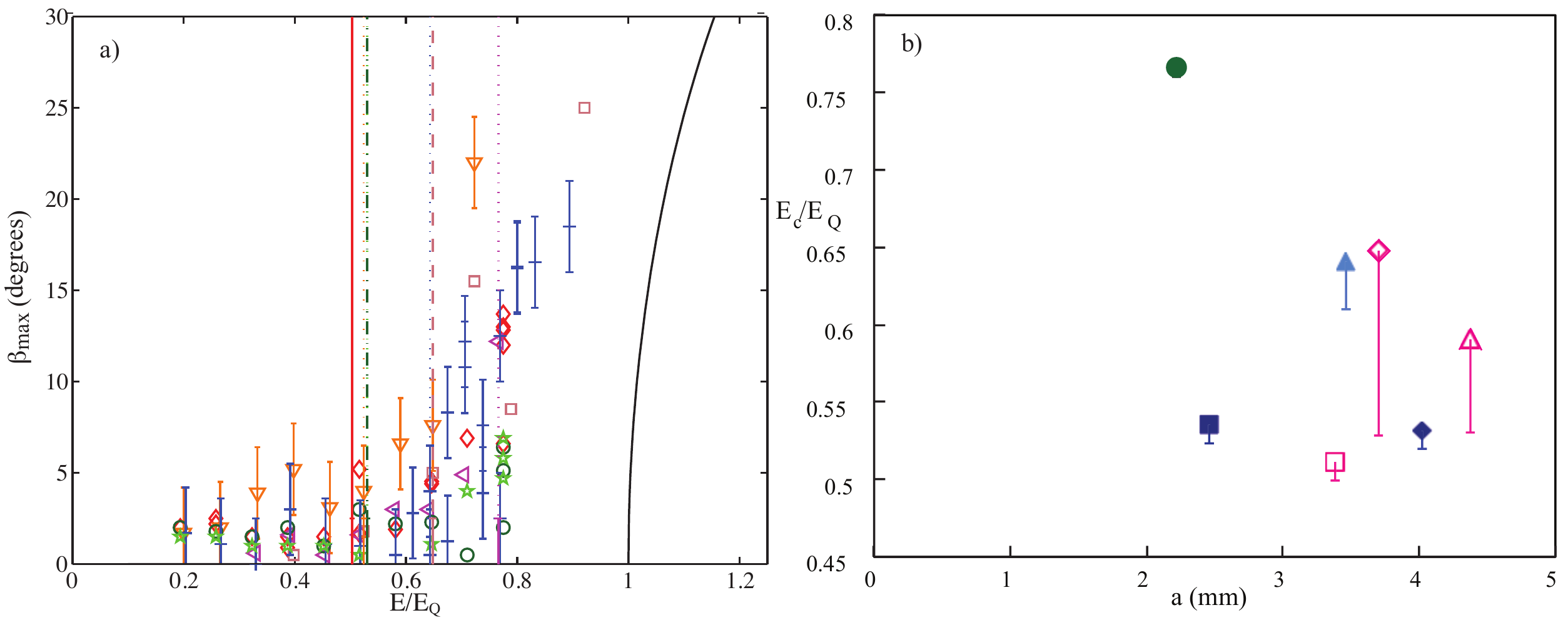}
\caption{\footnotesize{(a) Maximum inclination $\beta_{max}$ measured over 4 minutes as a function of $E/E_{Q}$}. The vertical lines represent the onset of particle sustained motion for the different drops, and the black thick line represents the prediction for the dipole orientation for a solid particle given by \refeq{quinckeAngle}. The symbols $\diamond$, $\square$, and $\bigtriangledown$ denote drops covered with Al ($r=12\mu m$) with  $\varphi = 93\%$, $88\%$, and $96\%$ and a = 1.7mm, 1.8mm, and 2.2mm, the symbol $\triangleleft$ corresponds to a drop covered with Pe ($\varphi = 78\%$, a = 2.2mm), the symbols $\circ$, and $\star$ denote drops covered with G ($r=8.5\mu m$) with  $\varphi = 100\%$ and $93\%$ and a = 1.2mm, and 2mm, 
the symbol $+$ corresponds to a drop covered with G ($r=3.5\mu m$) with $\varphi = 100\%$ and a = 1.7mm. { {For clarity,  error bars are shown only for some drops as they are similar for all the measurements. }}
(b) Electric field at which sustained drop tilt occurs  as a function of the drop diameter. The filled symbol $\circ$ represents a drop covered with Pe particles for 
$\varphi = 78\%$. The empty symbols represent drops covered with Al particles, $\diamond$ for $\varphi = 95\%$, $\square$ for $\varphi = 93\%$, and $\Delta$ for $\varphi = 88\%$. The filled $\Delta$ symbol represents a drop covered with glass sphere ($r = 3 \mu m$) for $\varphi = $. The filled $\square$ and $\diamond$ symbol represent respectively drops covered with glass spheres ($r = 12\mu m$) for $\varphi = 100\%$ and $\varphi = 94\%$. }
\label{fig:beta}
\end{figure}

   Figure \ref{fig:beta}.(a) shows the maximum inclination $\beta_{max}$ measured for drop covered with different particles as a function of $E/E_{Q}$.  The scatter of the data suggests  that the tilting effect is strongly dependent on particles characteristics. For comparison, the solid black line represents the theoretical dipole orientation prediction for a Quincke sphere \refeq{quinckeAngle} \cite{Salipante-Vlahovska:2010}.

   Figure \ref{fig:beta}.(b) summarizes  the threshold fields  as a function of the drop diameter for high surface coverage of different types of particles. Independently from the type of particle, the dimensionless critical field strength is in the range [0.5-0.8]. i.a. two to four times lower than the thresholds measured for ``clean'' drops and around twice lower than for a solid sphere.  Hence, high surface coverage of particles drastically reduces the critical fields strength for onset of  rotation.

\begin{figure}[h]
\includegraphics[width=3in]{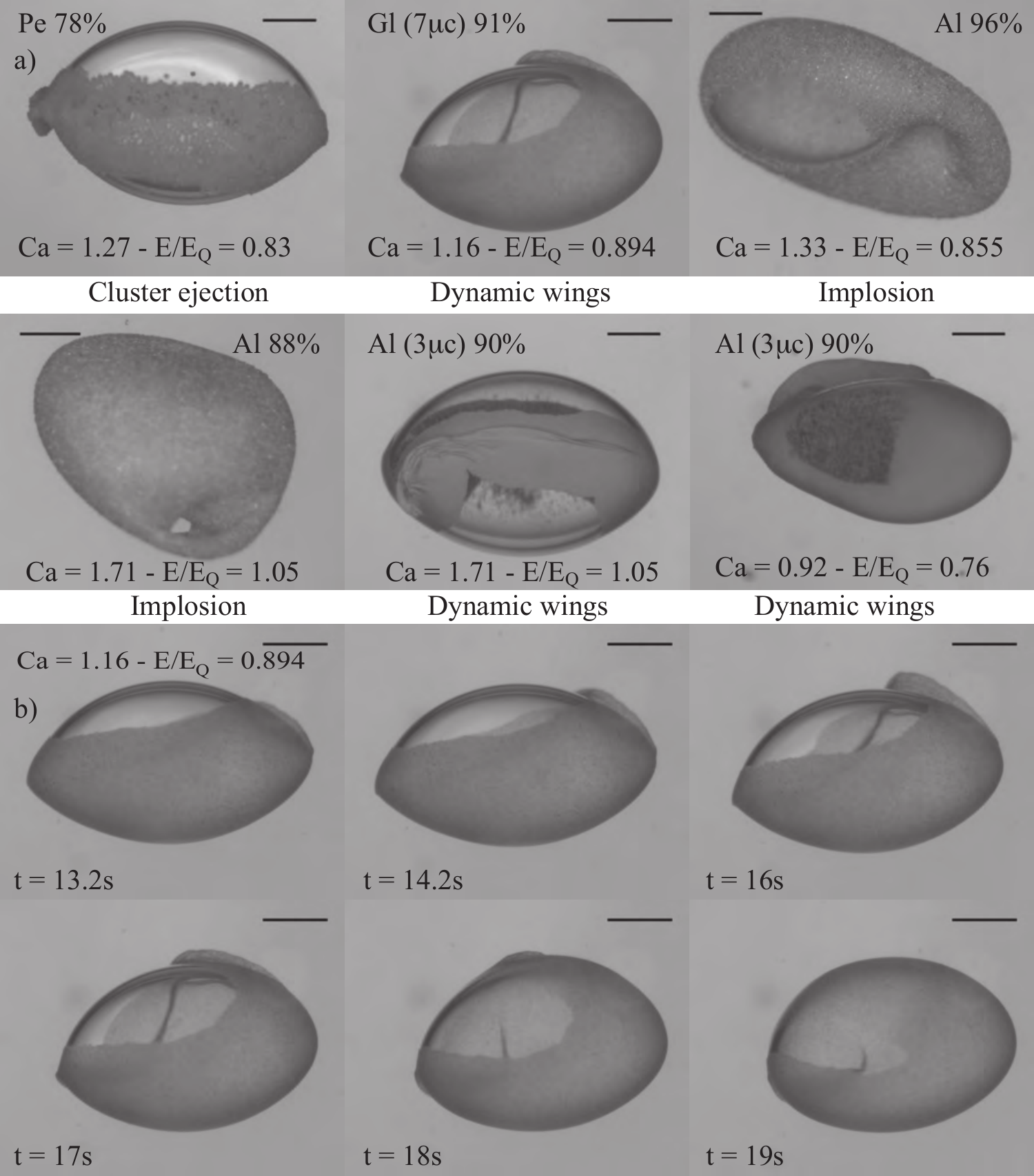}
\caption{\footnotesize{(a) Drop behaviors for very strong electrical fields. (b) Evolution of the ``dynamic wings". The scale bar corresponds to 1 mm.}}
\label{fig:DE}
\end{figure}

For very strong electrical fields, the drops exhibit very peculiar behavior, see  Figure \ref{fig:DE}.(a), as drop implosion  or bucking of the layer of particles leading to either ejection of cluster of particles similar to the ejection observed for lower surface coverage \cite{Ouriemi:2014}, or the formation of ephemeral dynamic wings.   Figure \ref{fig:DE}.(b) illustrates ephemeral wings formation.

\section{Discussion}
In summary, our experiments show that surface--adsorbed particles suppress  the electrohydrodynamic flow, enhance overall oblate drop deformation, and decrease the threshold for rotational flow.  
 Modification of the surface properties, e.g., surface tension, interfacial viscoelasticity and surface conductivity,    could be responsible for these effects. { {Here we discuss each of these mechanisms and show that the oblate deformation of the particle-covered drop is well captured by  a capsule model; the colloidal membrane endows the interface with shear elasticity which immobilizes the interface and suppresses the flow. The decrease in the tilt threshold, however,  can not be explained at this time with the available shell model \cite{Turcu:1989,Jones:1984}.}}
\subsection{Modification of the surface tension}
\label{subsec:surface_tension}

The increase of deformation could be related to a lower effective interfacial tension,  $\gamma_{eff}$, due to the particles. 
Fitting the data on  Figure \ref{fig4} with the Taylor model \refeq{Def_param}  yields  $\gamma_{eff}$ in the range [0.12 - 0.16] mN/m, which corresponds to a decrease of almost 97\% relative to the  tension for the clean interface. Impact of surface absorbed particles on the surface tension has been studied in relation with Pickering emulsion and liquid marbles\cite{Aussillous-Quere:01}. The presence of solid particles at the interface usually is associated  reduction of the liquid/liquid interfacial tension \cite{Nushtaeva:2004,Cengiz-Erbil:2013}. Theoretical modeling \cite{Levine:1989} shows that for dense monolayer of mono-disperse spherical particles, $\gamma_{eff}$ depends mainly on the contact angle of the particles, $\gamma_{eff} = \gamma [1-\pi (1-cos\theta)^2/4\sqrt{3}]$, and shouldn't decrease below 50\% of $\gamma$. This order of magnitude is confirmed by experimental measurement \cite{Vilkova:2013,Nushtaeva:2004}. For lower concentration of particles, almost no variation of $\gamma$ were measured \cite{Vignati:2003, Bormashenko:2013}. Recent publications  \cite{Cengiz-Erbil:2013, Bormashenko:2013} on marble liquids point out the strong dependence of $\gamma_{eff}$ on  size, concentration, surface free energy and water contact angle of the encapsulating powder. For non--densely packed particles, the capillary interactions between particles can be neglected, and the ``modified" surface tension can be expressed as  $\gamma_{eff} = \gamma + (\gamma_{sl}+\gamma_{sa}-\gamma)A_0/A$  \cite{Bormashenko:2013}, where $\gamma_{SL}$ and $\gamma_{SA}$ are respectively the surface tension at the powder/liquid and powder/air interface, and $A$ and $A_0$ are  the total surface of the marble and the surface covered by particles. Estimate for  all drops considered in our work show
that  $\gamma_{eff}$ should be relatively close to $\gamma$. Hence, a lower interfacial tension can not explain the strong increase of the drop deformation. Moreover, $\gamma_{eff}$ should vary with the electrical fields, as the packing changes with the electrical fields leading to a non linear variation with the deformation. In conclusion, a variation of the interfacial tension due to the presence of surface--adsorbed particles can not explain the strong deformation increase compared to a clean drop.

\subsection{Surface conductivity}
We can model the monolayer of surface-adsorbed particles as  a solid shell, with an  equivalent surface conductivity  estimated as the particle conductivity divided by the particle diameter (which is a measure of the thickness particle monolayer), $\sigma_s\sim \sigma_p/(2r)$. The current conservation boundary condition may be expressed as (assuming negligible charge convection, which is reasonable in weak fields and small drop deformations) $\sigma_\out E_n^\out-\sigma_\ins E_n^\ins=-\sigma_s \nabla_s \cdot \bE_t$.  The solution shows that the deformation is described by the same  discriminating function in the Taylor theory \refeq{Def_param}  but  with  $R$ replaced by  $R+2R_s$ where $R_s=\sigma_s a/ \sigma_\out$ is the dimensionless surface conductivity. This shows that oblate deformation should be suppressed as the particle conductivity increases and eventually the drop deformation should change from oblate to prolate. 
In our experiments however we observe enhancement of the oblate deformation by the surface-adsorbed particles, not suppression. This implies that surface conductivity is unlikely to play a significant role in drop deformation.

 \subsection{Interfacial viscosity}
 
 Particles at interfaces behave as two-dimensional suspension. Accordingly, the increased dissipation arising from the particle motions results in increased surface viscosities \cite{Fuller:review12, Lishchuk:2009, Lishchuk:2014b}. A drop with high surface viscosities effectively acts as a drop with very high bulk viscosity.  In the limit of high viscosity ratio Taylor's law gives
 \begin{equation}
D =\Ca \frac{9\left(-19+\left(5+9R+5R^2\right)S\right)}{80 S (2+R)^2}
\label{DMI}
\end{equation}
Increased surface viscosity also suppresses the surface flow, see \refeq{svel}.

\begin{figure}
 \centerline{\includegraphics[width=3in]{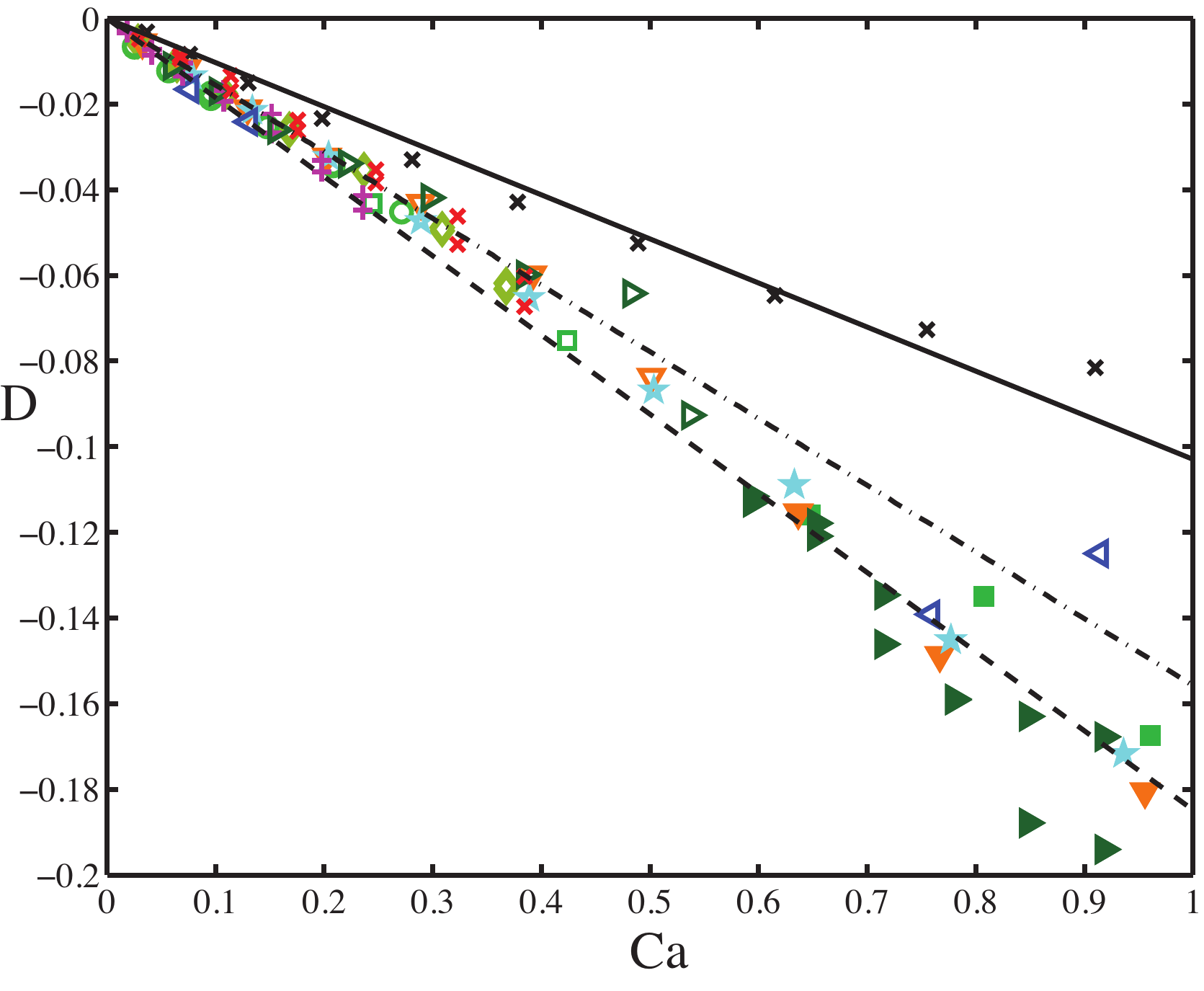}}
\caption{\normalsize{Drop deformation as a function of the capillary number. Symbols the same as in  Figure \ref{fig4}. The solid line represents the Taylor theory \refeq{Def_param}  for $\lambda = 0.07$ ,  the dashed line is \refeq{DMI} for $\lambda \rightarrow \infty $, and the lines for capsule model \refeq{Deqs} with shear elastic modulus $1.43 \gamma$.}}
\label{fig9}
\end{figure}

 Figure \ref{fig9} shows the comparison between the experimental data and the theoretical curve for an infinitely  viscous drop \refeq{DMI}. The theoretical curve is close to the data specially for low capillary numbers but still underestimates the measured deformation. Hence, the effect of the presence of particles can not be completely explained by the suppression of the electrohydrodynamic flow. Moreover, we also observe that drop deformation continues to increase even above the rotation threshold (see  Figure \ref{fig4}) in contrast to the theoretical prediction for high-viscosity drops \cite{He:2013} .

\subsection{Interfacial elasticity}
\label{subsec:capsule}

A monolayer of densely packed particles  at a liquid interface behaves similarly  to an elastic sheet \cite{Vella:2004, Pocivavsek:2008, Planchette:2013, Xu:2005, Datta:2010} and can support anisotropic stresses
and strains: it buckles in uniaxial compression and cracks under  tension. This solid-like behavior can be characterized in terms of Young's modulus and Poisson ratio derived from simple theoretical considerations see  \cite{Vella:2004,Cicuta:2009, Planchette:2013}.  
Recent theoretical model \cite{Erni:2012} shows that presence of a significant interfacial shear elasticity (greater than the interfacial tension) and of a nonzero interfacial compression elasticity can cause drops to buckle. { {In our experiments, the particle layer also eventually exhibit buckling, see  Figure \ref{fig:DE},suggesting that the particle-coated drop behaves  like a capsule (a drop enclosed by an elastic membrane).}}

The deformation of capsules in electric fields has been considered in only two studies \cite{Ha:2000c, Karyappa:2014}. Ha and Yang  \cite{Ha:2000c} developed a small deformation theory similar to the one for a capsule in external linear flows \cite{Barthes_Biesel-Sgaier:1985}.  
{ {Since the particles at the interface are closely  packed,  the monolayer of jammed particles can be much more easily sheared than compressed (compacted). Accordingly,  the particle monolayer can be modeled as an incompressible two-dimensional material and drop deformation is given by (see Appendix for details)}}
\begin{equation}
D_{eq} =\Ca_s\left[\frac{27((1+R)^2-4/S)}{32 (2+R)^2}\right]\quad \Ca_s=\frac{\eps_\out E_0^2 a}{G}=\Ca \frac{\gamma}{G}
\label{Deqs}
\end{equation}
where $G$ is the elastic shear modulus.   Figure \ref{fig9} show the data fit with $\gamma/G=0.7$. The value of $G$ is in reasonable agreement with value deduced from the Young's modulus of particle rafts \cite{Vella:2004}.

{ {\subsection{Threshold for drop tilt and electrorotation}

To estimate  the effects of the particle monolayer on the onset of electrorotation,  we consider the spherical shell \cite{Turcu:1989,Jones:1984}.  We can model the particle-covered drop as a layered sphere  with effective properties $\eps_p$ and $\sigm_p$.
%\frac{a}{2r}\left(\eps_p +\sigma_p/(\im \omega)\right)$. 
For a thin shell \cite{Jones:2003}
\begin{equation}
\label{eqShell}
\eps_s=\frac{\eps_p \eps_\ins }{\eps_p+\delta \eps_\ins}\,,\quad \sigm_s=\frac{\sigm_p \sigm_\ins }{\sigm_p+\delta \sigm_\ins}\,,\quad\delta=\frac{2r}{a} \ll 1
\end{equation}
where the shell thickness corresponds to the particle diameter $2r$.
%permittivity $\eps_\ins$ and conductivity $\sigm_\ins$ (corresponding to the drop fluid) coated with a thin layer with thickness $2r$ (corresponding to the particle diameter) characterized by $\eps_p$ and $\sigm_p$.
While $S$ is less affected by the particles, $R$ can be significantly changed due to the wide range of particle conductivities. 
For example, \refeq{eqShell} shows that presence of a resistive (very low-conductivity) layer decreases $R$ and  thereby lowering the rotation threshold \refeq{quinckeW}. However, since this decrease is also observed for conductive particles, surface conduction most likely not a relevant explanation. 
%The decrease is independent of particle conductivity and can not be explained by shell model \cite{Turcu:1989,Jones:1984} . 
A  possible explanation may be  drop asphericity; prolate rigid ellipsoids do have lower threshold for electrorotation than spheres
 \cite{Cebers:2000, Dolinsky-Elperin:2009}. This issue requires further examination both experimentally and theoretically.}}
 
\section{Conclusions}
\label{sec:conc}

We experimentally study the effect of high concentration of surface--adsorbed particles on drop deformation  in a uniform DC electrical field. The fluid system consists of a silicon oil drop suspended in castor oil, both very weakly conducting liquids. A broad range of  particle sizes, conductivities, and shapes is explored. 
In weak electric fields, the presence of particles enhances the deformation compared to a particle--free (clean)  drop and suppresses the electrohydrodynamic flow. Drop deformation is well described by a capsule model, which treats the  particle monolayer as an elastic sheet.  In stronger fields, drops tilt due to the Quincke effect  but the onset  is significantly  lowered compared to the clean drop. The decrease can not be explained with the existing spherical shell model and likely due to drop asphericity.  Even stronger electrical fields give rise to  more exotic behaviors as ephemeral dynamic wings or drops implosion. 
The similarity between the particle-covered drop and a capsule provides promising new insight into the impact of particles on interfacial dynamics. Our findings open questions ranging from understanding the fluid/solid transition that occur for high surface coverage to stability of Pickering emulsions. We hope our work will stimulate further research on the electrohydrodynamics of particles at interfaces.

\section{Acknowledgement}
This research  was supported by NSF awards CBET-1132614 and CBET- 1437545.

\appendix
{ {
\section{Small-deformation theory for an initially spherical elastic capsule placed in a uniform DC electric field}
An uniform DC electric field, $\bE=E_0 \zhat$, exerts electric pressure and shear  on a sphere with non-capacitive interface (continuous electric potential)\cite{Taylor:1966}
%deforms into an ellipsoid because of the electric traction exerted 
\begin{equation}
\label{eqE}
\bt^\el=p^\el \left(1+3 \cos2\theta\right)\rhat+\tau^\el_s \sin 2\theta {\bm{\hat\theta}}
\end{equation}
where 
%the pre tangential electric traction is
\begin{equation}
p^\el=-\frac{3 \left(\Rr^2+1-2/\Sr\right)}{4 (\Rr+2)^2}\,,
\quad \tau^s=\frac{9 (\Rr-1/\Sr)}{2(\Rr+2)^2}\,.
\end{equation}
The electric tractions deforms the interface; the material particles at the interface move to new position described by
\begin{equation}
\label{perturbation of shape}
\bx_s=a\left[\left(1+s \left(1+3 \cos2\theta\right)\right)\rhat+ u \sin 2\theta {\bm{\hat\theta}}\right]
%\,,\quad \bX=\bx_p(t)+\left(1+g(\theta, \varphi, t)\right)\rhat\,,
\end{equation}
 The in-plane displacement does not generate overall shape change but creates elastic stresses that oppose the electric shear and immobilize the interface. The deviation from sphericity is qualified by $s$ 
% due to displacement of material points
%The displacement of a material point, $\bd=\eps \left(f(\theta, \varphi, t) \rhat+\bd_\tang \right)$, is decomposed into a radial and tangential to a sphere components.
\begin{equation}
%r_s=a\left[1+s \left(1+3 \cos2\theta\right)\right]\,,\quad 
D=3 s\,,\quad d_{||}=r_s(0)=1+4s  \,,d_\perp=r_s(\pi/2)=1-2s
\end{equation}
%where $\eps$ quantifies the deviation from sphericity, e.g., capillary number. 
For small deformations the membrane behaves as an elastic-Hookean material and the elastic stresses are \cite{Ha:2000c, Karyappa:2014, Vlahovska_swing:2011}
\begin{equation}
\label{eqM}
\bt^m=p^m \left(1+3 \cos2\theta\right)\rhat+\tau^m_s \sin 2\theta {\bm{\hat\theta}}\,,
\end{equation}
where
\begin{equation}
\tau^m =2 \left(2 \Ca_s^{-1} u+3 \Ca_a^{-1}(2 s+u)\right)\,\quad p^m=2 \Ca_a^{-1}(2 s+u)\,.
\end{equation}
$\Ca_s$ and $\Ca_a$ are the dimensionless shear and extensional elasticities; $\Ca_s=\eps_\out E_0^2 a/G$.
%Note that &if the membrane is area--inextensible $u=2s$ 
%and $p^m$ vanishes
Balancing the electric and elastic stresses, \refeq{eqE} and \refeq{eqM}, yields
\begin{equation}
s=\Ca_s\frac{\tau^\el_s}{8}-\left(3\Ca_s+2 \Ca_a\right)\frac{p^\el}{8}
\end{equation}
In the limit of inextensible membrane, $\Ca_a\rightarrow 0$, we obtain \refeq{Deqs}.
}}

%\bibliographystyle{unsrt}
%\bibliography{refs2015}
\providecommand*\mcitethebibliography{\thebibliography}
\csname @ifundefined\endcsname{endmcitethebibliography}
  {\let\endmcitethebibliography\endthebibliography}{}

\end{document}